\documentclass[pra,aps,twocolumn,showpacs,floatfix]{revtex4}
\usepackage{mathrsfs}
\usepackage{rotating}
\usepackage{color}
\usepackage{graphicx}           
\usepackage{dcolumn}            
\usepackage{bm}                 
\usepackage{hyperref}           
\usepackage[latin1]{inputenc}   
\usepackage{pstricks}           
\usepackage{amsmath,amssymb}
\usepackage{version}
\usepackage{epstopdf}
\DeclareGraphicsExtensions{.eps,.ps}

\usepackage{ulem}
\usepackage{float}
\usepackage{booktabs}
\usepackage{subfigure}
\usepackage{graphicx}
\usepackage{dcolumn}
\usepackage{bm}

\begin{document}

\preprint{APS/123-QED}
\title{Longitudinal photoelectron momentum shifts induced by absorbing a single
XUV photon in diatomic molecules}
\author{Di Lao}
\author{Pei-Lun He}
 \email{a225633@sjtu.edu.cn}
\author{Feng He}
 \email{fhe@sjtu.edu.cn}
\affiliation{%
Key Laboratory for Laser Plasmas (Ministry of Education) and Department of
 Physics and Astronomy, Collaborative Innovation Center of IFSA
 (CICIFSA), Shanghai Jiao Tong University, Shanghai 200240, China
}%

\date{\today}
\begin{abstract}
The photoelectron momentum shifts along the laser propagation are investigated by
the time-dependent perturbation theory for diatomic molecules, such as H$_2^+$,
N$_2$ and O$_2$. Such longitudinal momentum shifts characterize the photon
momentum sharing in atoms and molecules, and oscillate with
respect to photon energies, presenting the double-slit interference structure.
The atomic and molecular contributions are disentangled analytically, which gives
intuitive picture how the double-slit interference structure is formed.
Calculation results show the longitudinal photoelectron momentum distribution
depends on the internuclear distance, molecular orientation and photon energy. The
current laser technology is ready to approve these theoretical predictions.
\end{abstract}

\maketitle

\section{\label{sec:level1}Introduction}
The rapid development of laser technology has enabled the discovery of
many novel phenomena appearing in laser atoms/molecules
interactions, among which ionization is one of the most fundamental processes \cite{krausz00}. Many ultrafast measurements
are based on ionization and subsequently induced
processes \cite{krausz09}. In ionization, photon energies, as well as photon momenta
are absorbed by molecular or atomic systems from laser fields.

The concept about ionization has been developed from Einstein's photoelectric
effect, to multiphoton ionization, above threshold ionization, and tunneling
ionization \cite{becker}. In all these processes, photoelectrons carry photon energies, and gain
momenta mainly in the laser polarization plane. In these studies, the dipole
approximation are widely accepted when a Ti:Sapphire laser pulse with
an intensity below 10$^{16}$ W/cm$^2$ is introduced \cite{Forre06}. Within the dipole
approximation, the photoelectron momentum distribution along the laser
propagation direction has a symmetric distribution centered at
zero (see Ref \cite{PhysRevLett.114.103004,PhysRevA.90.013418,PhysRevLett.105.133002} for example).

However, due to the small magnitude of a laser wave vector $\bf {k}$,
the transferred momentum is obscured. This situation
justifies the widely adopted dipole approximation in atomic
physics, where $\lvert \bf {k} \rvert$ is set to be
zero. Due to the fact that the dipole approximation is expected to
work well when the
wavelength of the laser is much longer than the target size,
almost all previous investigations beyond the dipole approximation
were using short wavelengths and
focused on nondipole asymmetry \cite{Krassig95,Jung96,Hemmers01,Ricz03,Bilheux03,Hemmers06,Grum14}. In those
topics beyond the dipole approximation \cite{Reiss13,Klaiber13,Yakaboylu13,Reiss214,Ivanov15,Chelkowski15}, the law of transferred momentum is one of the most
interesting and important topics to study \cite{Smeenk11,Titi12,Liu13,Ludwig14,Chelkowski14,Cricchio15,Krajewska15}. Also, the transferred
momentum due to photo-ionization processes is interpreted as a
significant part of radiation pressure, which is of astronomer's
interest \cite{Michaud70}.

It was only very recently that the partition of absorbed photon
momenta between nuclei and electrons has been addressed \cite{Smeenk11}.
It was found \cite{Chelkowski14} that for circularly polarized
laser pulse in the tunnelling regime, the law of partition is $\langle p_z^e
\rangle=\frac {\langle E_k \rangle}{c}+\frac {0.3I_p}{c}$, $\langle
p_z^i \rangle=\frac {0.7I_p}{c}$, where $E_k$ is the photoelectron
energy, $I_p$ is the ionization potential, $c$ is the light speed, $\langle p_z^{e,i}
\rangle$ is the expectation value of longitudinal electron or ion momentum.
For a linearly polarized light,
situations are complex due to the coulomb interaction between nuclei
and recoiled electrons \cite{Liu13,Ludwig14}.
While in the single photon limit,
the transferred momentum of electrons and nuclei can be expressed as
$\langle p_z^e \rangle=\frac {8}{5}\frac {E_k}{c}$, $\langle p_z^i
\rangle=\frac {8}{5} \frac {I_p}{c}-\frac {3}{5} \frac
	      {\omega}{c}$ when the
electron is initially in the 1s state \cite{Chelkowski14}.

Energy sharing between electrons and nuclei has been studied in the
  laser-molecule interactions \cite{Esry12,Wu13,Silva13}, while momentum
  sharing has not been addressed in molecules.
In this paper, we studied the longitudinal photoelectron momentum in
  diatomic molecules
in the single photon ionization regime by the time-dependent perturbation theory.
A double-slit interference pattern \cite{Cohen66,Kushawaha13,XJLiu14}for the longitudinal photoelectron
  momentum distribution is reported,
and the interference patterns in H$_2^+$, N$_2$ and O$_2$ are compared
  and analyzed in details. The rest of this paper is organized as
  following. In Sec.II we introduce the numerical models. The
  calculation results for H$_2^+$, N$_2$ and O$_2$ are presented in
  Sec. III. We end the paper in Sec. IV with a short conclusion.

\section{Numerical Methods}
The single-photon ionization of H$_2^+$ in XUV fields can be studied
by the time-dependent perturbation theory, which expresses the
transition amplitude as
\begin{eqnarray}
M(\textbf{p})=-i\int\!\mathrm{d} t\langle \psi_{f}(t) | H_{I} |\psi_{0}(t)\rangle,
\label{transition}
\end{eqnarray}
where the initial state $ \langle \textbf{r}|\psi_0(t)\rangle=\psi_0(\textbf{r})\exp[-iI_p
  (t-t_0)] $ with $t_0$ being the
  starting time of the interaction, the final state is described by a
  plane wave  $\langle \textbf{r}|\psi_f(t)\rangle =\langle \textbf{r}|\textbf{p}(t)\rangle=\exp(i{\textbf{p}
  \cdot \textbf{r}}) \exp[-i\frac{p^2}{2}(t-t_0)]$.
The interacting Hamiltonian $H_I$ is
\begin{equation}
H_{I}={\bf A}(t,z)\cdot {\bf p}+ \frac {1}{2} {{\bf A}(t,z)}^2
\end{equation}
where ${\bf A}(t,z)$ is the laser vector potential. We use atomic
units throughout this paper unless indicated otherwise. We consider the laser
electric field propagates along $+\hat z$ direction, and its polarization
axis is in the $x-y$ plane. Thus, the vector potential is
\begin{equation}
\textbf{A}(t,z)=\frac {A_0}{\sqrt{1+\epsilon ^2}} \left [ { \cos(\omega t-kz) \hat x } {+{\epsilon} \sin(\omega t-kz)
  \hat y } \right ],
\end{equation}
where $k$ is the wave number or the photon momentum.
When $\epsilon=0$ or 1, the laser field is linearly or circularly
polarized. For an infinite long laser pulse, the integration in
Eq. (\ref{transition}) yields
\begin{eqnarray}
M({\bf p} )\propto \frac{ A_{0}}{\sqrt{1+\epsilon^2}}(p_{x}+i\varepsilon p_{y}) \langle {\bf p}
-k \hat {z} \! \mid \! \psi_{0} \rangle \delta (\omega-p^2/2-I_{p}).
\end{eqnarray}
The $\delta (\omega-p^2/2-I_{p}) $ guarantees the energy conservation.
Finally, the expectation value of the photoelectron
momentum $p_z$ can be calculated via
\begin{eqnarray}
\langle p_{z} \rangle = \frac {\oint_{S} \! \mathrm{d}^3 {\bf p} \,
  p_{z} (p_{x}^{2}+\varepsilon^2 p_{y}^{2}) \! \mid \! \langle {\bf
    p}-k \hat {z} \! \mid \! \psi_{0} \rangle \! \mid ^{2}}{\oint_{S}
  \! \mathrm{d}^3 {\bf p} \, (p_{x}^{2}+\varepsilon^2 p_{y}^{2})\!
  \mid \! \langle {\bf p}-k \hat {z} \! \mid \! \psi_{0} \rangle \!
  \mid ^{2}},
\label{eq:nine}
\end{eqnarray}
where $S$ represents the integral surface satisfying $\frac {1}{2}p^2=\omega-I_{p}$.
$|\langle {\bf p}-k \hat {z} \! \mid \! \psi_{0} \rangle|^2$ is the initial momentum
probability distribution for the bound electron after shifting $k\hat z$.

For H$_2^+$, the molecular orbitals can be roughly
constructed by combining the two atomic states, i.e.,
\begin{equation}
\psi_{g/u}(\textbf{r})=\psi_{\textrm{atom}}(\textbf{r}-\textbf{R}/2)\pm \psi_{\textrm{atom}}(\textbf{r}+\textbf{R}/2),
\end{equation}
where $|\textbf{R}|$ is the internuclear distance and $\psi_{\textrm{atom}}$
is the atomic state. The corresponding molecular wavefunction in momentum
representation is
\begin{eqnarray}
|\psi_{g}(\textbf{p})|^2&=&|\psi_{\textrm{atom}}(\textbf{p})|^2\cos^2(\textbf{p}\cdot \textbf{R}/2),\nonumber \\
|\psi_{u}(\textbf{p})|^2&=&|\psi_{\textrm{atom}}(\textbf{p})|^2\sin^2(\textbf{p}\cdot \textbf{R}/2). \label{momentum}
\end{eqnarray}
Insertion of Eq. (\ref{momentum}) into Eq. (\ref{eq:nine}) yields the
the expectation value of $p_z$,
\begin{equation}
\langle p_z \rangle=\langle p_z \rangle_\textrm{atom}+\langle p_z \rangle_\textrm{osc}, \label{two}
\end{equation}
which are contributed by the atoms and the interference of two
centers, respectively.
The conclusion indicated by Eq. (\ref{two}) was deduced from H$_2^+$,
but it should work for general diatomic molecules.
For more complex diatomic molecules, such as N$_2$ and O$_2$, we
calculate the molecular orbitals by using the MOLPRO \cite{molpro}. In later
calculations for N$_2$ and O$_2$, we set the bond lengths at
$R_{\textrm{N}_2}=2.07\, a.u.$ and $R_{\textrm{O}_2}=2.2\, a.u.$,
respectively. In MOLPRO, we used the Gaussian type orbital (GTO)
basis cc-pVTZ and calculated the coefficients and exponents for
different GTO basis, with which the molecular orbitals are constructed.

\section{Calculation Results}
\begin{figure}
\centering
\includegraphics[width=0.5\textwidth]{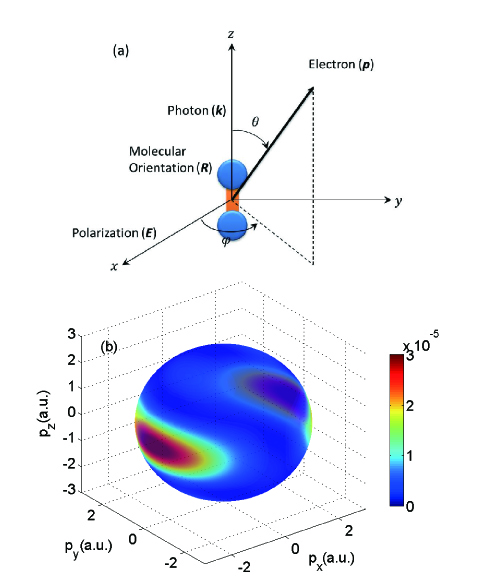}
\caption{\label{fig1}(a) Geometry applicable to a linearly polarized
  field with H$_2^+$ aligned along $z$ axis. (b) The photoelectron
  momentum angular distribution of H$_2^+$ with internuclear distance
  $R=2$ and photon energy $\omega=4.5$ a.u..}
\end{figure}

For atoms initially in different states, the transferred longitudinal
momenta from photon momenta are different \cite{Seaton95}. For molecules,
we expect the longitudinal momenta are more complex due to multi-coulombic centers.
For H$_2^+$ at $R=2$ a.u., we set $\lambda=1.236$ and the molecular
wavefunction is written as
\begin{equation}
 \psi_0({\bf r})=\frac {\lambda^{3/2}}{\sqrt {\pi}}(e^{-\lambda \mid {\bf r}-\frac {1}{2}
  {\bf R} \mid}+e^{-\lambda \mid {\bf r}+\frac{1}{2} {\bf R} \mid}).
\label{ground}
\end{equation}
The laser-molecule
interaction geometry is sketched in Fig. \ref{fig1} (a).
The XUV field propagates along $+z$ axis, and its polarization axis is in the $x-y$ plane.
The photoelectron momentum angular distribution is sketched in Fig. \ref{fig1} (b).

Within the dipole approximation, the photoelectron only gains momenta in the laser
polarization plane from the laser field.
The momentum distribution along the laser propagation axis
is symmetric with respect to $p_z=0$.
Therefore the expectation value of longitudinal momentum $\langle p_z\rangle$ should be 0.
In the laser polarization plane, the Coulomb potential drags the photoelectron when it escapes from
the parent ion, and
gives rise to a tilt angle for the photoelectron angular distribution \cite{He15}.
Though the Coulomb potential modifies the photoelectron distribution, it does not change
the fine structures. Especially, when the electron escapes from the nucleus very quickly,
the Coulomb action can be neglected. In the following calculations, we use very
high-energetic photons, thus the Coulomb potential can be safely neglected.
It has also been shown in Ref \cite{Chelkowski14} that the distribution of $p_z$ and $\langle p_z \rangle$ are
not affected by the Coulomb corrections when the electron is initially
in 1s atomic state .

After introducing the nondipole effect, the center of the longitudinal momentum
is shifted away from $p_z=0$ to $p_z=k$.
According to Eq. (\ref{two}), $\langle p_z \rangle$ of diatomic molecules
may present more complex structures beyond the shift which happens in
atoms. For H$_2^+$, when the electron
is kicked by the photon along $+z$ axis, the electron may fly away
from both nuclei, thus $\langle p_z \rangle$ component may show some
interference patterns. To numerically prove that, we insert Eq. (\ref{ground})
into Eq. (\ref{eq:nine}) and obtain
\begin{equation}\begin{split}
&\langle p_z \rangle =\\
&\frac {\int_{S} \! \mathrm{d}^3 {\bf p} \, p_{z}
      (p_{x}^{2}+p_{y}^{2}) \frac{1}{(\lambda^2+({\bf p}-k\hat
	z)^2)^{4}} \cos^2 [\frac {\bf R}{2}\cdot ({\bf p}-k \hat
      z)]}{\int_{S} \! \mathrm{d}^3 {\bf p} \, (p_{x}^{2}+p_{y}^{2})
      \frac{1}{(\lambda^2+({\bf p}-k\hat z)^2)^{4}} \cos^2 [\frac {\bf
	R}{2}\cdot ({\bf p}-k \hat z)]} ,
\end{split}\label{slits}
\end{equation}
The cosine term in the integration in Eq. (\ref{slits}) carries the
double-slit interference, which should depend on both the molecular
orientation and internuclear distance.

\begin{figure}
\centering
\includegraphics[width=9cm]{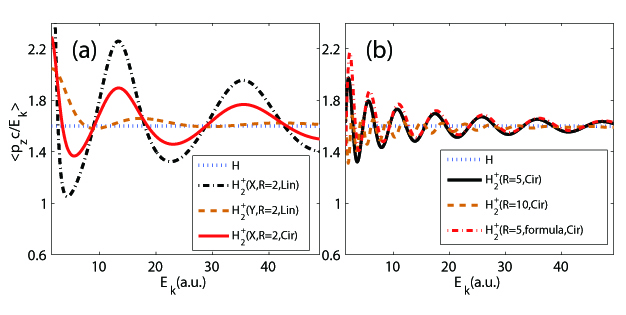}
\caption{\label{fig2} (a) The expectation value of the longitudinal photoelectron momentum
  as a function of the photoelectron kinetic energy.
The blue dotted line represents the result for a hydrogen atom, and
  the black dash-dotted line, brown dashed line, red solid line are
  the results for H$_2^+$ aligned along
  $x$ axis with a linearly polarized laser pulse,
  aligned along $y$ axis with a linearly polarized laser pulse, and
  aligned along $x$ axis with a circularly polarized laser pulse. The
  internuclear distance is 2 a.u..
  (b) The expectation value of longitudinal photoelectron momentum with respect to
the photoelectron energy for $R=5$ (black solid line) and
  $R=10$ (brown dashed line) is shown. The blue dotted line is same as
  that in (a). The red dash-dotted line is the analytical result
  governed by Eq. (\ref{analytic}). The circularly polarized XUV field
  is implemented.}
\end{figure}

Fig. \ref{fig2} (a) shows the expectation value of longitudinal momentum $\langle p_z\rangle$
as a function of the photoelectron energy. The black dash-dotted curve and the
brown dashed curve are for H$_2^+$ aligned along $x$ axis and $y$ axis,
respectively, and the XUV field is linearly polarized along the $x$ axis. The
red solid curve is for the case that H$_2^+$ is aligned along $x$ axis and
the XUV field is circularly polarized in $x-y$ plane. The dotted horizontal line
indicates $\langle p_z \rangle$ for a hydrogen atom in the ground state. It is clear
that $\langle p_z \rangle$ oscillates around the equilibrium position
$\langle p_z \rangle=1.6 E_k/c$. The oscillation amplitude gradually decays with the
increasing of the photoelectron energy. And the oscillation amplitude depends on the
orientations of molecule. When the molecular axis and the laser
polarization axis are parallel to each other, the oscillation amplitude
is larger than that when these two directions are orthogonal to each other. When the laser
field is circularly polarized, the oscillated $\langle p_z \rangle$ is similar to
the case using linearly polarized laser pulse after averaging over all molecular
orientations. The oscillation of $\langle p_z \rangle$ also depends on the
internuclear distance, as shown in Fig. \ref{fig2} (b). For a larger
internuclear distance, the separation between neighboring peaks is smaller,
which is consistent to the general double-slit interference pattern.

The fluctuation of $\langle p_z \rangle$ can be
viewed in an analytical form.
In the high-photon-energy limit, $\frac{1}{(\lambda^2+p^2+k^2-2kp\cos
  \theta)^4}$ in Eq. (\ref{slits})
can be further expanded as $\frac{1}{(\lambda^2+p^2+k^2)^4 }(1+\frac{8kp\cos
\theta}{\lambda^2+p^2+k^2})$ by discarding high-order terms.
With this, when the molecular axis is parallel to the laser propagation axis,
Eq. (\ref{slits}) can be analytically written as
\begin{equation}
\langle p_z \rangle = \frac {8}{5} \frac {E_k}{c}\biggl[ 1 - \frac
  {6}{R^2 E_k}\cos \alpha \cos \beta -\frac {15c}{8\sqrt{2}R^2
    E_k^{3/2}} \sin \alpha \sin \beta \biggr],
\label{analytic}
\end{equation}
where $\alpha=Rp$, $\beta=Rk$.
It is clear that the first term in Eq. (\ref{analytic})
is the atomic contribution, and the two latter terms lead to the
oscillation of $\langle p_z \rangle$. In high energy limit, the second term is more
important than the third term since the third one decays
faster. $\cos(\alpha)$ clearly describes the double-slit interference
for the photoelectron releasing from two nuclei.
$\cos(\beta)$ describes the double-slit interference contributed by
the photon momentum. The product of $\cos(\alpha)$ and
$\cos(\beta)$ contributes to the main oscillation of $\langle p_z
\rangle$ when $E_k$ is large.
We plotted $\langle p_z \rangle$ governed by Eq. (\ref{analytic}) in Fig. \ref{fig2}
(b) for $R=5$ a.u. One
may clearly see that Eq. (\ref{analytic}) matches the
simulation results very well especially when $E_k$ is very large.

\begin{figure}
\centering
\includegraphics[width=9.0cm]{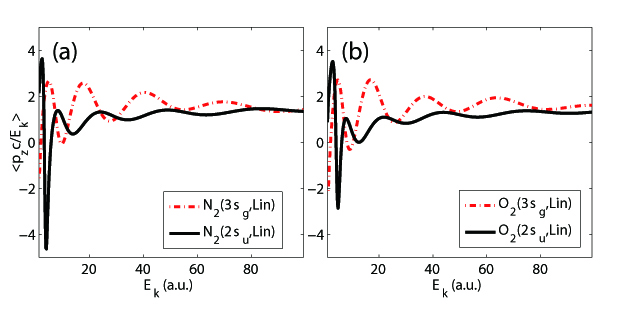}
\caption{\label{fig3}
(a) The expectation value of the longitudinal momentum distribution as a function of $E_k$
for the photoelectron initially in $3\sigma_g$ (red dash-dotted curve) and
$2\sigma_u$ (black solid curve) of N$_2$. The molecule is aligned along $z$
axis, and the XUV filed is linear polarized along $x$ axis. (b) Same
as (a) but for O$_2$. The ionization potentials for $2\sigma_u$ and $3\sigma_g$ of
N$_2$ (O$_2$) are 0.78 (1.08) and 0.63 (0.745) a.u., respectively.
}
\end{figure}

The double-slit interference showing in $\langle p_z \rangle$ exists not only in the
simplest molecule H$_2^+$, but also in more general diatomic
molecules. According to Eq. (\ref{eq:nine}), the ultimate $\langle p_z \rangle$ should
also depend on initial molecular orbitals.
We now study the $\langle p_z \rangle$ of the photoelectron from N$_2$
and O$_2$. Figure \ref{fig3} (a) shows the $\langle p_z \rangle$ as a
function of E$_k$ for the photoelectron initially in $3\sigma_g$
(red dash-dotted curve) and $2\sigma_u$ (black solid curve) states. The
$3\sigma_g$ orbital is constructed by two symmetrical atomic $2p_z$
states, and the $2\sigma_u$
orbital is constructed by two asymmetrical atomic $2s$ states.
We only consider the response of a
single electron in the XUV fields though electrons in the inner orbital
might have larger cross section to be ionized. This assumption will
capture some phenomena qualitatively and work as a prototype. The double-slit
interference pattern is still observed. This oscillated $\langle p_z \rangle$ finally
converges to atomic case with the increasing of $E_k$. The similar
behavior are preserved for O$_2$, as shown in Fig. \ref{fig3}
(b). In both panels, the phases of the oscillated
$\langle p_z \rangle$ from $2\sigma_u$ and $3\sigma_g$ are opposite, which is due to the
opposite phase for these two orbitals.

\begin{figure}
\centering
\includegraphics[width=9.0cm]{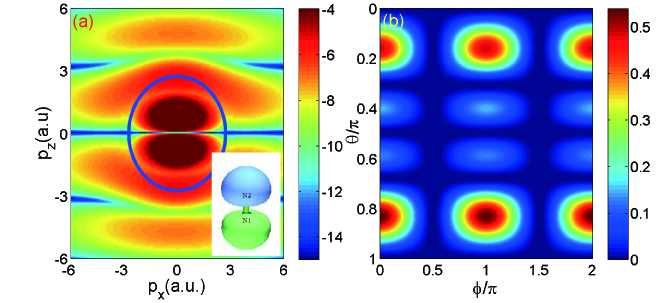}
\caption{\label{fig4} (a) The momentum probability distribution for
  the bound electron in $2\sigma_u$ of N$_2$. (b) The photoelectron momentum
angular distribution for the electron initially at $2\sigma_u$  of
  N$_2$.
 }
\end{figure}

When $E_k$ is relatively small, $\langle p_z \rangle$ could be negative though the laser
propagates along $+ z$ axis. And as a matter of fact, this can be understood by looking into the atomic states.
For $2\sigma_u$ of N$_2$, which is constructed by two asymmetrical atomic 2s states, the transferred longitudinal momenta
depend on the formula $\frac{8}{5} \frac{E_k}{c}(1-\frac{1}{E_k})$ \cite{Seaton95}. Viewing from this formula, negative value of
$\langle p_z \rangle$ occurs when $E_k$ is relatively small. Alternatively, the negative
value of $\langle p_z\rangle$ can be understood by looking into Eq. (\ref{eq:nine}),
which shows the photoelectron momentum distribution is proportional to
${ |\langle p-k \hat z|\psi_0\rangle|^2}$.
Fig.~\ref{fig4}(a) shows ${|\langle p-k \hat z|\psi_0\rangle|^2}$ for $2\sigma_u$ of
N$_2$. The sketched molecular orbital in space coordinate is
shown in the right-bottom corner. The momentum probability distribution for
the bound electron is symmetric with respect to $p_z=k$ and is asymmetric
with respect to $p_z=0$ though $k$ is only slightly different from
zero. $\delta (\omega-p^2/2-I_p)$ manifests
itself as a ring satisfying ${ p_x^2+p_z^2=2(\omega-I_p)}$ in the
plane $p_y=0$, as shown
by the circle in Fig. \ref{fig4} (a) \cite{He215}. The non-negligible $k_z$ results
in the upward shift of the electron momentum distribution before it is
ionized, which makes the probability in the ring not symmetric any
longer in the upper and lower half spaces. When $E_k=3.72$, the radius
of the ring is 2.73 a.u.
Actually, the local maximum in the lower half space moves closer to the
ring, on the contrary, the local maximum in the upper half space moves away from
the ring, thus the photoelectron momentum with $p_z<0$ has larger
probabilities, though it is not clear in the logarithmic scale.
The calculated photoelectron angular distribution is shown in Fig. \ref{fig4} (b).
The asymmetric distributions in the upper and lower half spaces make the the averaged
$p_z<0$. For a different photon energy $\omega$, the ring in Fig. \ref{fig4} (a) will
meet other local maximum of minimum, thus the photoelectron probability in
the upper half space may be larger than that in the lower half space. Therefore,
$\langle p_z\rangle$ oscillates with respect to $\omega$ or $E_k$. Of course, in the dipole
approximation, $\langle p_z\rangle$ is always zero because the small
$k_z$ is neglected, and thus the symmetric distribution with respect to $p_z=0$ is
always preserved.

\section{Conclusions}

In conclusion, by including the photon momentum transferred into the
atom beyond the dipole approximation, the expectation value of longitudinal photoelectron
momentum $\langle p_z\rangle$ shifts away from zero. In diatomic molecules, $\langle p_z\rangle$
oscillates with respect to the photon energy. Two factors contribute
to such oscillation: the double-slit interference of the
photoelectron $\cos(\alpha)$ and the double-slit interference
of the photon $\cos(\beta)$. As
shown above, the interference pattern occurs in all diatomic
molecules. The present work indicates the importance of the photon
momentum sharing in photoionization. The fruitful structures of
$\langle p_z\rangle$ offers another perspective to extract molecular information.

\section*{acknowledgements}
This work was supported by NSF of China (Grant No. 11104180, 11175120, 11121504, 11322438),
and the Fok Ying-Tong Education Foundation for Young Teachers in the
Higher Education Institutions of China (Grant No. 131010 ).

\end{document}